\documentclass[fleqn,usenatbib,onecolumn]{mnras}

\usepackage{newtxtext,newtxmath}

\usepackage[T1]{fontenc}

\DeclareRobustCommand{\VAN}[3]{#2}
\let\VANthebibliography\thebibliography
\def\thebibliography{\DeclareRobustCommand{\VAN}[3]{##3}\VANthebibliography}

\usepackage{graphicx}	
\usepackage{amsmath}	
\usepackage{tabularx}
\usepackage{epsfig}
\usepackage{supertabular}
\usepackage{lscape}
\usepackage{babel}
\title{Pushchino Multibeam Pulsar Search. X. Observations of pulsars at declinations above $+53^\circ$}

\author[S. A. Tyul'bashev et al.]{
	S. A. Tyul'bashev, $^{1}$\thanks{E-mail: serg@prao.ru}
	G.E. Tyul'basheva,$^{2}$
	M.A. Kitaeva $^{1}$
	\\
	$^{1}$ Pushchino Radio Astronomy Observatory, Astro Space Center, Lebedev Physical Institute, Russian Academy of Sciences, Pushchino, 142290 Russia \\
	serg@prao.ru\\
	$^{2}$ Institute of Mathematical Problems of Biology RAS (IMPB RAS) brunch of Keldysh Institute of Applied Mathematics of Russian Academy of Sciences, \\
	Pushchino, Moscow reg., Russia \\
}

\date{ }

\pubyear{ }

\begin{document}
	\label{firstpage}
	\pagerange{\pageref{firstpage}--\pageref{lastpage}}
	\maketitle
	
\begin{abstract}
A search for pulsars was carried out using a Large Phased Array (LPA) radio telescope at a frequency of 110.4 MHz with a time resolution of 3.072 ms and a frequency resolution of 19.5 kHz with a 2.5 MHz bandwidth used. The survey was conducted in a site with declinations of $+53^{\circ}<\delta<+87^{\circ}$. The viewing area is approximately 4100 sq.deg. The search was carried out using Fourier power spectra. To increase sensitivity, multiple observations were made in each direction in the sky, and the resulting power spectra were summarized. This made it possible to increase sensitivity by about 5-10 times, depending on the direction in the sky. A blind search opened 35 known pulsars. Estimates of the flux density for 33 pulsars have been obtained.

Keywords: pulsar search, low frequency observations, flux density, average profile	
\end{abstract}

	\section {Introduction}

As is known, pulsars were discovered in 1967 by \citeauthor{Hewish1968}, \citeyear{Hewish1968}. In the well-known pulsar catalog ATNF \footnote{https://www.atnf.csiro.au/people/pulsar/psrcat/} at November 2025 there are more than 4300 pulsars listed in \citeauthor{Manchester2005}, \citeyear{Manchester2005}, of which about 2000 have been discovered in the last 10 years (see, for example, \citeauthor{Sanidas2019}, \citeyear{Sanidas2019}; \citeauthor{Cameron2020}, \citeyear{Cameron2020}; \citeauthor{Han2021}, \citeyear{Han2021}; \citeauthor{Tyulbashev2022}, \citeyear{Tyulbashev2022}; \citeauthor{Zhou2023}, \citeyear{Zhou2023}; \citeauthor{Han2025}, \citeyear{Han2025} and others). The motivation for these surveys is to search for new types of pulsars, to search for pulsars with extreme properties, to study the interstellar medium, to study the properties of pulsar samples, and more.

For example, the use of new search methods led to the discovery of rotating radio transients (RRATs), which are pulsars that radiate rare sporadic pulses (\citeauthor{McLaughlin2006}, \citeyear{McLaughlin2006}). The search for pulsed dispersed emission led to the discovery of fast radio bursts (FRBs) (\citeauthor{Lorimer2007}, \citeyear{Lorimer2007}). Long-period transients with periods of tens of minutes have been discovered by Caleb et al. and Hurley-Walker et al. (\citeauthor{Caleb2022}, \citeyear{Caleb2022}; \citeauthor{HurleyWalker2022}, \citeyear{HurleyWalker2022}). Their nature is unclear, and they may be either radio pulsars or white dwarfs (\citeauthor{Rea2024}, \citeyear{Rea2024}). These recent discoveries highlight the importance of conducting searches to understand the nature of sources with pulsed radio emission.

Pulsar searches are one of the main tasks for new high-end antennas using broadband recorders. So, the search for pulsars at FAST has led to the discovery of about 1000 pulsars in a few years \footnote{http://zmtt.bao.ac.cn/GPPS/; http://groups.bao.ac.cn/ism/CRAFTS/202203/t20220310{\_}683697.html} (\citeauthor{Wang2019}, \citeyear{Wang2019}; \citeauthor{Han2021}, \citeyear{Han2021}). A small number of new pulsars have been discovered on the CHIME telescope by Good et al. (\citeauthor{Good2021}, \citeyear{Good2021}), but the telescope has proven to be an outstanding tool for searching for FRBs \footnote{https://www.chime-frb.ca/catalog/} by Amiri et al. (\citeauthor{Amiri2021}, \citeyear{Amiri2021}).

Regular searches for pulsars may also be associated with the emergence of new techniques, using which new objects are discovered in archived data. For example, when processing archived data using machine learning and a new way to speed up counting, 23 new pulsars were discovered by \citeauthor{Morello2019}, \citeyear{Morello2019}. Using neural networks, it was possible to discover 4 RRATs based on data where a search for pulsed radiation had already been conducted by \citeauthor{Tyulbashev2022a}, \citeyear{Tyulbashev2022a}. Using the fast folding algorithm (FFA), two pulsars were discovered in the archived data of GMRT by \citeauthor{Singh2022}, \citeyear{Singh2022}.

Pushchino multibeam pulsars search (PUMPS; \citeauthor{Tyulbashev2022}, \citeyear{Tyulbashev2022}) is carried out on the Large Phased Array (LPA) radio telescope (\citeauthor{Shishov2016}, \citeyear{Shishov2016}; \citeauthor{Tyulbashev2016}, \citeyear{Tyulbashev2016}). After the modernization of the telescope, two independent antenna pattern appeared. On one of them (the LPA3 radio telescope), 128 beams are formed simultaneously, which allows to view half of the celestial sphere (declinations $-9^{\circ}<\delta<+55^{\circ}$) in a 24-hour (monitoring) mode. The developed search method uses data accumulated over 10 years, which allows for a tenfold increase in sensitivity when searching for pulsars (\citeauthor{Tyulbashev2022}, \citeyear{Tyulbashev2022}). This led to the discovery of about 200 new pulsars and RRATs \footnote{ https://bsa-analytics.prao.ru/en/project/about/ }.

The second antenna pattern (LPA1 radio telescope) allows for the formation of no more than 8 beams simultaneously. In total, there are 416 beams in this system that overlap declinations from $-16^{\circ}$ to $+86^{\circ}$. Thus, some declinations available in LPA1 are not available for LPA3. In this paper, we consider the first search results on the LPA1 radio telescope at declinations of $\delta > +53^{\circ}$.

\section{Observations and processing}\label{sec:observations}

	\begin{figure}
		\includegraphics[width=0.99\columnwidth]{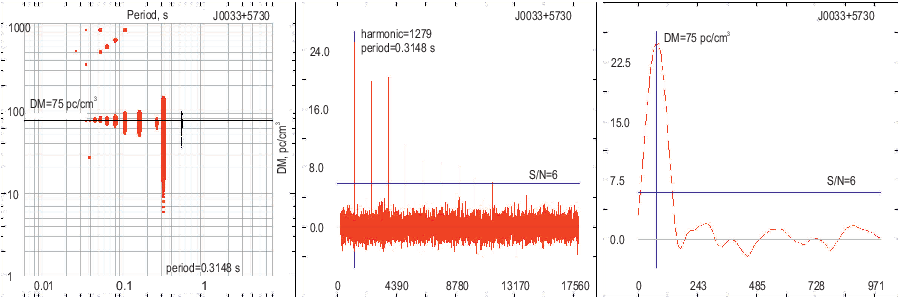}
		\caption{The left panel shows a $P/DM$ map of the famous pulsar J0033+5730. The intersection of the horizontal and vertical lines indicates the probable value of $P$ and $DM$. The middle panel shows the power spectrum, which corresponds to the point of intersection of the lines. The right panel shows the harmonic height in the power spectrum in units of S/N, depending on the $DM$ being tested. The maximum in this dependence corresponds to the $DM$ expected for the found pulsar. On the horizontal and vertical axes of the figures from left to right: period and dispersion measure; the number of the point in the power spectrum and the harmonic height in conventional units; the dispersion measure and signal-to-noise.}
		
		\label{Fig1}
		\end{figure}

The LPA radio telescope at the Lebedev Physical Institute is a meridian instrument. The LPA diagramming system creates beams that are located in the plane of the meridian. Therefore, it is possible to record a source that falls within the declinations covered by the LPA beams only once per day. The beam size is approximately $0.5^{\circ} \times 1^{\circ}$, which corresponds to a transit time through the meridian of approximately 3.5 minutes at half power. As noted in the Introduction, this paper discusses the search for pulsars on the LPA1 radio telescope. Unlike LPA3, which has a beam intersection at a power level of 0.405, LPA1 has a higher beam density. The beams are overlapped at a level of 0.8. As a result, LPA1 does not have significant sensitivity variations due to the source being off-center in the beam pattern.

The central reception frequency is $\nu=110.4$~MHz. The reception bandwidth of LPA1 is the same as that of LPA3, and is equal to 2.5~MHz. The pulsar recorder used for the survey allows for a wide range of changes in the point sampling time and the number of frequency channels within the bandwidth. We selected a point sampling time of 3.072~ms and a channel width of 19.5~kHz (128-channel reception mode).

Narrow frequency channels allow for compensation of dispersion smoothing within the channel, and for almost all dispersion measures ($DM$), the pulse broadening due to dispersion smoothing is less than the pulse broadening due to scattering. Therefore, there is no additional sensitivity loss associated with smoothing.

Observations at LPA1 were started in April 2019 and have been ongoing for 6 years. LPA1 is the main instrument of the PRAO, and it is used for various observations. On average, 8-9 days per month are allocated for conducting a survey. During these observations, a band of approximately $3^{\circ}$ in declinations is covered, but with the intersection of the beams at the level of 0.405. To overlap the beams at the level of 0.8, additional observations are conducted in the beams that are located in the middle between the main beams. Observations are conducted in hourly portions. Time control is carried out using the GPS system. For some of the observation beams, it has not yet been possible to conduct observations, so some known strong pulsars are not included in the final table.

The survey on LPA1 is made for declinations above $+53^{\circ}$, i.e. there was a small overlap of declinations available for LPA3. The LPA1 radio telescope also has the possibility of observing declinations in the interval $-16^{\circ}<\delta<-9^{\circ}$, which are also not available on LPA3. However, at low declinations, the effective area of the LPA decreases in proportion to the cosine of the zenith distance, and there is also a very high level of interference from large cities located about 100 km to the north (Moscow) and to the south (Tula) of the LPA location (Pushchino). Therefore, no survey of pulsars for negative declinations was conducted at the LPA1.

Taking into account the observations in the main and intermediate beams, a strip with a declination width of 3$^\circ$ can be viewed in two days. Declinations from $+53^{\circ}$ to $+87^{\circ}$ can be covered in 23 days. On average, each direction should have been viewed for 21 days over a period of 5 years, but during the survey, several times more time was allocated for observations of the Galactic plane than for directions outside the Galactic plane. This is because the background temperature in the Galactic plane is 2 or more times higher than outside the Galactic plane, according to \citeauthor{Turtle1962}, \citeyear{Turtle1962}, and observations in the Galactic plane must be conducted for 4 times longer to achieve comparable sensitivity.

The sensitivity of the LPA in the search for pulsars is considered in the work \citeauthor{Tyulbashev2016}, \citeyear{Tyulbashev2016}, where it is shown that outside/in the Galaxy plane, with a single observation, it is possible to see slow pulsars ($P \sim 1$~c) with an integral flux density of 6-8 and 15-20~mJy. The sensitivity of the search decreases as the cosine of the source height. However, all the directions under study are close to the zenith, and the maximum drop in sensitivity due to this factor is less than 0.85. The intersection of the LPA1 beams is implemented at a level of 0.8, which also results in a slight loss of sensitivity for pulsars whose coordinates fall between the beams.

On the other hand, the time of crossing the meridian depends on the declination of the source, and if for pulsars located at the equator, this time is 3.5 minutes, then at declination $60^{\circ}$ it is $3.5/\cos(60^{\circ})=7$~minutes. Increasing the observation time by a factor of 2 leads to an increase in sensitivity by a factor of $2^{1/2}=1.4$. At the highest declinations, the duration of an observation session can increase by a factor of 10, and the sensitivity of a single observation session can increase by a factor of about 3.

To increase the sensitivity, we used the summing of power spectra. If all the characteristics are stable and change little, the increase in sensitivity should be proportional to the square root of the number of added sessions. As shown in the paper \citeauthor{Tyulbashev2020}, \citeyear{Tyulbashev2020}, due to interference and the slightly changing sensitivity of the radio telescope, the overall sensitivity loss can reach 20-30\%.

To process the observations, we modified the program that was used to work with the LPA3 data. The detailed work of the pulsar search program based on the summed power spectra is described in the paper (\citeauthor{Tyulbashev2022}, \citeyear{Tyulbashev2022}). The modification of the program was that, in addition to adding the power spectra for different days, the power spectra obtained during a single session were also added if the duration of the source's passage through the meridian was 2, 3, or more standard durations of 65536=$2^{16}$ points (201.33~s $\approx$ 3.4 minutes). A standard duration refers to the time it takes for the source to pass through the meridian along half of the LPA beam pattern at a declination of $\delta =0^\circ$. That is, at a declination of $\delta =60^\circ$, the time of passage through the meridian will be two standard durations, and so on.

As a result, the search using the summed power spectra allowed to increase the sensitivity in the survey for pulsars out of the Galactic plane by about 3-4 times compared to a single session. In the Galactic plane, the expected increase in sensitivity is 5-6 times, but in absolute values it is 20-30\% lower than out of the Galactic plane. Thus, the maximum expected loss of sensitivity is $0.9 \times 0.85 \times 0.7=0.54$ (1.85 times), and the gain in sensitivity due to the accumulation of power spectra and the extension of observation time is 5-10 times.

For each direction in the sky, the expected sensitivity can be calculated. In this survey, the sensitivity is at least 1.5-3~mJy for any direction with $DM<100$~pc/cm$^3$ and $P>0.5$~s. Recently, a survey was conducted to search for pulsars covering the area $+53^{\circ}<\delta<+87^{\circ}$ using the LOFAR aperture synthesis system (\citeauthor{Haarlem2013}, \citeyear{Haarlem2013}) at a central receiving frequency of 135 MHz. The best sensitivity in the LOTAAS survey reached 1.2~mJy, but for pulsars with small periods and $DM>100$~pc/cm$^3$, it degrades to 5~mJy \citeauthor{Sanidas2019}, \citeyear{Sanidas2019}. Taking into account the steep spectra of pulsars (\citeauthor{Maron2000}, \citeyear{Maron2000}), where the typical value of the spectral index is $\alpha = 1.8 \, (S \sim \nu^{-\alpha})$, the best sensitivity is 1.7-7.2~mJy at the LPA reception frequency. Thus, the sensitivities in the reviewed works were close.

Pulsars were searched visually using $P$ vs. $DM$ maps, which display cumulative power spectra, and signal-to-noise ratio (S/N) vs. $DM$ in the average profile. In blind searches, we do not know the $DM$ or the width of the average profile ($W$) of a candidate for a new pulsar. Therefore, it is necessary to obtain Fourier power spectra by assuming different $DM$ and pulse width for the candidates. The search for possible $DM$ values was conducted in the range of 0-1000 pc/cm$^3$. Of course, searching for pulsars at $DM>200-300$ pc/cm$^3$ may be unnecessary for the meter range. However, the power spectrum is not calculated for all $DM$ values, as it is necessary to take into account the broadening of pulses within frequency channels and due to scattering. This allows for a variable step size based on dispersion (see the methodology in \citeauthor{Tyulbashev2022}, \citeyear{Tyulbashev2022}) and the actual calculation of 125 different dispersion measures. We may have chosen an overly high $DM$ limit, but we hoped that if a candidate had a period longer than a few seconds, there would be a chance to detect it at $DM>300$ pc/cm$^3$ in the second, third, and other harmonics.

The second parameter to be varied is related to the width of the average profile. It is clear that when averaging wide profiles, the observed S/N profile will increase. We varied the averaging from 3.072~ms to 196.608~ms. In this case, 3.072~ms is the count of power spectra without averaging the raw data, $2 \times 3.072=6.144$~ms is the count with averaging over 2 points, and so on, doubling the step until averaging $64 \times 3.072=196.608$~ms.
\begin{figure*}
	\includegraphics[width=0.99\columnwidth]{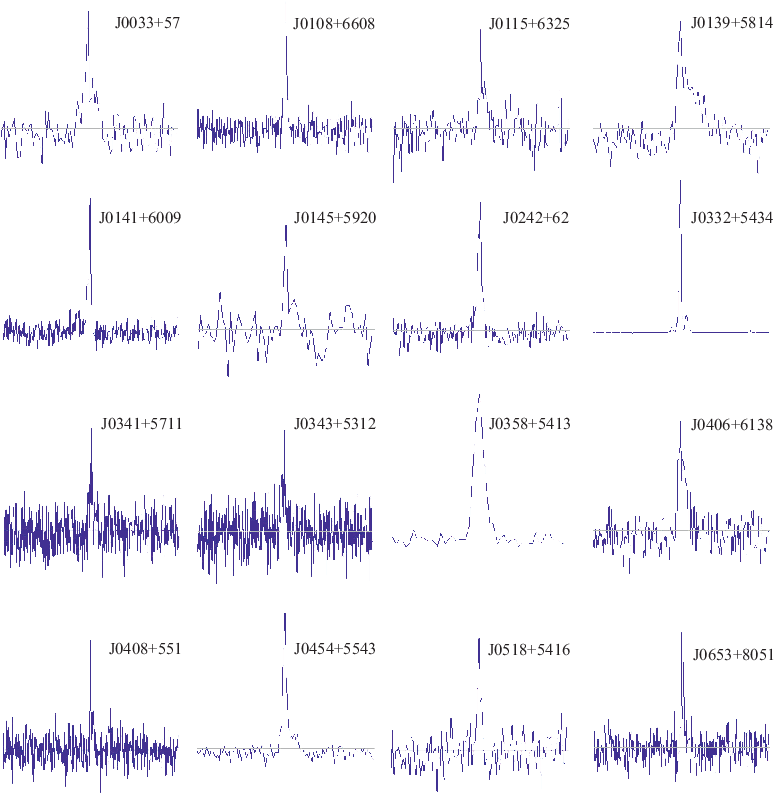}
	\caption{Average profiles of known pulsars detected by blind search at declinations of $+53^{\circ}<\delta<+87^{\circ}$.}
	\label{Fig2}
\end{figure*}
\begin{figure}
	\includegraphics[width=0.99\columnwidth]{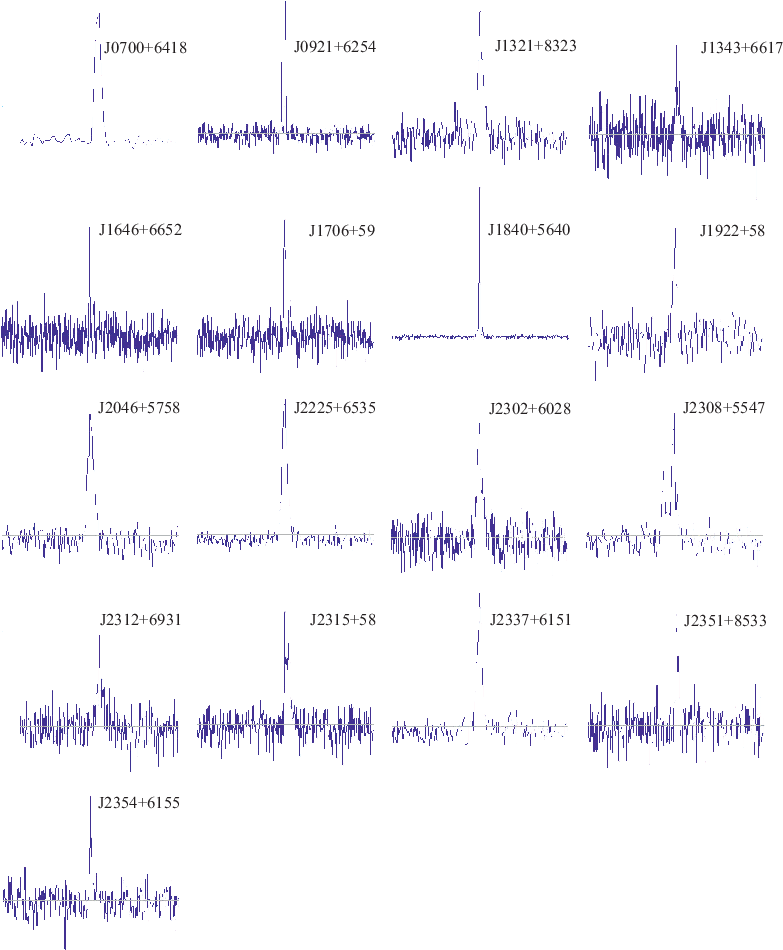}
	\contcaption{}
	\label{Fig2}
\end{figure}

In total, we calculate power spectra by iterating over 125 different $DM$ and 8 assumed pulse widths. As a result, 1000 ($125\times 8$) averaged power spectra are created for each direction in the sky. The $P/DM$ map simultaneously displays all the obtained spectra, which show the harmonic frequencies in the power spectra that exceed a specified level. Fig. \ref{Fig1} shows an example of the analyzed images for the well-known pulsar J0033+5730.

After obtaining an estimate of the pulsar's $P$ and $DM$ values from the summed power spectrum, its characteristics are refined. To do this, an average profile is searched for at values close to the estimated $P$ and $DM$ values. It is assumed that the maximum value of the S/N in the average profile at certain verified values of $P$ and $DM$ corresponds to the refined values of the $P$ and $DM$. It is not always possible to obtain an average profile, as the pulsar is often not detected in individual sessions lasting several minutes, even though it is clearly visible in the summed power spectrum.

The peak flux density was estimated based on the observed S/N ratio in the average profile and the estimated background noise level in the pulsar direction. The observed noise signal standard deviation ($\sigma_n$) in the recording is determined by the system temperature ($T_{sys}$), the effective antenna area ($A_{eff}$), the observation bandwidth ($\Delta \nu$), the recording time ($\tau$), and the number of averaged pulses ($N$). We use the radiometric gain formula:

$$
	\sigma_n = \frac{2 k T_{sys}}{A_{eff} (\tau  \Delta \nu  N)^{1/2}}
$$

For any pulsar, $\sigma_n$ is a constant determined by the background temperature in the pulsar's direction and the pulsar's period. In this case, the observed value of the S/N is $A/\sigma_n$, where $A$ is the amplitude value in the average profile and $\sigma_n$ is the standard deviation of the noise. The system temperature is determined by the sum of the ambient and background temperatures. For the ambient temperature, we assume a constant value of $T=290$~K. The background temperature ($T_b$) is determined from the isophotes obtained at a frequency of 178 MHz and converted to a frequency of 110.4 MHz using the law $T_b \sim \nu^{-2.55}$ (\citeauthor{Turtle1962}, \citeyear{Turtle1962}). Thus, the system temperature is $T_{sys}=T + T_b$. The integral flux density ($S_{int}$) is determined by integrating the intensity values of the average profile points divided by the number of points in the profile.

\section{Results}\label{sec:results}

When processing the observations on the $P/DM$ maps, 35 known pulsars were detected. Table 1 shows the extracted parameters for 33 of the 35 pulsars. For two pulsars, comments are provided below. The first column shows the name of the pulsar as it is presented in the ATNF. Columns 2-5 display the $P$ and $DM$ values of the pulsars from the ATNF catalog and our observations. The majority of sources in Table 1 are strong pulsars that have 10 or more days of observations, and therefore the $P$ and $DM$ estimates are obtained by averaging the best observational sessions. The LPA reception band is narrow, so the accuracy of $DM$ determination is low. We provide our estimate of $DM$ for comparison with the ATNF values. Estimates of $P$ show that values obtained from the LPA differ from the catalog values by a factor of $\sim 10^{-4}-10^{-5}$c. The asterisk in the second column indicates the low accuracy of the period in the ATNF. The values of the refined periods agree with the values obtained from the LOFAR observations (\citeauthor{Sanidas2019}, \citeyear{Sanidas2019}). The contents of columns 2 and 3 for ATNF and columns 4 and 5 for LPA data allow you to see the accuracy of the extracted estimates. Columns 6-7 show the peak and integral flux densities. Column 8 shows the half-width of the average profile. The half-width refers to the full width of the profile at half its height. Columns 9-10 show the flux density values at frequencies of 135 MHz and 102.5 MHz obtained from the LOFAR (\citeauthor{Sanidas2019}, \citeyear{Sanidas2019}) and LPA (\citeauthor{Malofeev2000}, \citeyear{Malofeev2000}) radio telescope before its reconstruction.

We estimate the accuracy of determining the peak and integral flux density (columns 6 and 7) to be 50\% of the given value. This low accuracy in estimating the flux is due to the fact that we have not controlled several factors that affect the accuracy. The most significant factors are the potential shift in the coordinate due to ionospheric scintillations, which can lead to incorrect corrections due to the pulsar's lack of alignment with the beam, as well as the potential for refractive scintillations in pulsars.

\begin{table}
\caption{Parameters of known ATNF pulsars found in a blind search}
\begin{tabular}{|l|c|c|c|c|c|c|c|c|c|}
	\hline
Name & $P_{ATNF}$ & $DM_{ATNF}$ & $P_{A}$ & $DM_{A}$ & $S_p$ & $S_{int}$  & $W_{50}$  & $S_{135}$ & $S_{102}$\\
    & (s) & (pc/cm$^3$) & (s) & (pc/cm$^3$) & (Jy) & (mJy)  & (ms)  & (mJy) & (mJy) \\
\hline
J0033+57         &    0.315*     & 75.6 & 0.31454   & $75\pm1.5$    & 0.44 & 19   & 9.5   & 29.7 & 24\\
J0108+6608       &    1.28365    & 30.5 & 1.28358   & $29 \pm 1.5$  & 2.7  & 47   & 20.7  & 39.2 & 23\\
J0115+6325       &    0.52145    & 65   & 0.52148   & $66\pm1.5 $   & 0.5  & 10   & 16.0  &  --  & --\\
J0139+5814       &    0.27245    & 73.8 & 0.27247   & $73.5\pm1.5$  & 0.66 & 53   & 17.6  & 197.6& 181\\
J0141+6009       &    1.22294    & 34.9 & 1.2229    & $35.1\pm0.5$  & 2.4  & 55   & 21.5 &  --   & --\\
J0147+5922       &    0.19632    & 40.1 & 0.19633   & $40\pm 2$     & 0.29 & 9.1  & 6.1  &  26.3 & 25\\ 
J0242+62         &    0.592*     & 3.92 & 0.5917    & $3.6\pm0.5$   & 1.1  & 31   & 12.0 &  48.7 & --\\
J0332+5434       &    0.71451    & 26.7 & 1.41458   & $26.5\pm0.5$  & 27.6 & 400  & 8.2  &  --   & 517\\
J0341+5711       &    1.888*     & 101  & 1.8875    & $103\pm3.0 $  & 0.74 & 11   & 55.0 &  6.7  & --\\
J0343+5312       &    1.93447    & 67.3 & 1.9346    & $68.3\pm2.0$  & 0.77 & 11   & 26.6 &  --   & 50\\
J0358+5413       &    0.15638    & 57.1 & 0.156395  & $57.4\pm0.5$  & 1.27 & 80   & 10.0 &  88.4 & --\\
J0406+6138       &    0.59457    & 65.4 & 0.59455   & $64.0\pm1.5$  & 0.66 & 23.9 & 19.6 &  55.2 & 19.6\\
J0408+551        &    1.837*     & 55   & 1.8377    & $55.3\pm0.5$  & 1.08 & 12   & 15.4 &  6.6  & --\\
J0454+5543       &    0.34072    & 14.5 & 0.34076   & $14.3\pm0.5$  & 2.55 & 77   & 7.9  &  65.9 & 150\\
J0518+5416       &    0.34020    & 42.3 & 0.34023   & $42.7\pm1.5$  & 0.29 & 10.0 & 11.0 &  11.2 & --\\
J0653+8051       &    1.21444    & 33.3 & 1.2142    & $31.8\pm1.5$  & 0.74 & 7.0  & 15.6 &  13.1 & 16\\
J0700+6418       &    0.19567    & 8.77 & 0.1957    & $8.7\pm0.5 $  & 0.98 & 31   & 5.6  &  --   & --\\
J0921+6254       &    1.56799    & 13.1 & 1.56801   & $13.2\pm0.5$  & 1.6  & 14.5 & 8.0  &  10.8 & 30\\
J1321+8323       &    0.67003    & 13.3 & 0.67004   & $13.7\pm1.0$  & 0.58 & 15.7 & 17.5 &  25   & 36 \\
J1343+6634       &    1.39410    & 30.0 & 1.39411   & $30.8\pm 1.0$ & 0.3  & 4.3  & 23.0 &  --   & --\\
J1647+6608       &    1.55979    & 22.5 & 1.59982   & $22.8\pm 1.0$ & 0.6  & 3.6  & 6.5  & 7.8   & --\\
J1706+59         &    1.47699    & 30.8 & 1.47649   & $29.7\pm1.5$  & 0.73 & 5.9  & 8.4  & 15.7  & --\\
J1840+5640       &    1.65286    & 26.7 & 1.65278   & $26.5\pm0.5$  & 13.5 & 111  & 10.2 &  55   & 50\\
J1922+58         &    0.52962    & 53.7 & 0.52962   & $55.0\pm1.5$  & 0.37 & 7.3  & 9.8  &  8.8  & --\\
J2046+5708       &    0.47673    & 101.7& 0.47673   & $101\pm3  $   & 0.73 & 26.6 & 16.5 & 18.5  & 34 \\
J2225+6535       &    0.68254    & 36.4 & 0.68254   & $35.2\pm1.5$  & 1.7  & 43.6 & 16.6 & 126.3 & --\\
J2302+6028       &    1.20640    & 156.7& 1.20643   & $157\pm3  $   & 0.92 & 24.2 & 27.0 & 68.5  & --\\
J2308+5547       &    0.47506    & 46.5 & 0.47508   & $46.3\pm1.5$  & 1.05 & 29.8 & 9.2  & 133.4 & --\\
J2312+6931       &    0.81337    & 71.6 & 0.81332   & $71.6\pm1.5$  & 0.86 & 15.7 &13.8  & --    & --\\
J2315+58/+5617   &    1.061*     & 73.2 & 1.0616    & $72.5\pm2.0$  & 0.93 & 21.1 &28.2  & 29.0  & --\\
J2337+6151       &    0.49536    & 58.4 & 0.49546   & $57.7\pm1.5$  & 1.37 & 41.5 &12.6  & 28.7  & 75 \\
J2351+8533       &    1.01172    & 38.2 & 1.0117    & $38.5\pm1.5$  & 0.33 & 5.5  &13.0  & 2.4   & --\\
J2354+6155       &    0.94478    & 94.6 & 0.94478   & $94.0\pm2.0$  & 0.84 & 15.4 &13.2  &  --   & --\\
		\hline
	\multicolumn{10}{|p{12cm}|}{\footnotesize\it Note: Pulsar J2315+58 is published in \citeauthor{Hessels2008}, \citeyear{Hessels2008}, where its declination is indicated as $\delta=58^{\circ}$. The angular resolution of the GBT radio telescope at 350 MHz is comparable to the angular resolution of the LPA at 111 MHz and is approximately $0.7^{\circ}$. We have discovered a pulsar that has $P$ and $DM$ close to pulsar J2315+58, but its declination according to our data is $56^{\circ}17'$, which is beyond the errors of the coordinates of the survey made on GBT.}\\
	\hline		
\end{tabular}
	\label{tab:1}
\end{table}

A comparison of columns 2,3 and 4,5 shows that the accuracy of $DM$ determination is generally better than 1.5~pc/cm$^3$, and the accuracy of $P$ is generally better than $10^{-4}$~s. It should also be noted that there were virtually no estimates of the flux density at 110.4 MHz prior to this work.

In the work \citeauthor{Bilous2016}, \citeyear{Bilous2016} the flux densities of 158 pulsars were collected and their integral spectra were constructed. Our estimates of the flux density for pulsars J0033+57, J0108+6608, J0139+5814, J0406+6138, J0454+5543, J0653+8051, J0700+6418, J0921+6254, J1321+8343, J1840+5640, J2046+5708, J2225+6535 and J2308+5547 confirm the cut offs of spectra or their flattening. Comparison of the integral flux densities at a frequency of 110.4 MHz (present work) and at 135 MHz (\citeauthor{Sanidas2019}, \citeyear{Sanidas2019}) pulsars (J0147+5922, J0242+62, J0358+5413, J0518+5416, J1647+6608, J1706+59, J1922+58, J2302+6028, J2315+58) is not included in the work \citeauthor{Bilous2016}, \citeyear{Bilous2016}, also indicate possible flattening or cut offs in the integral spectra. Of the 33 pulsars included in Table 1, 13 (40\%) pulsars have a cut off or flattening of the spectrum shown in \citeauthor{Bilous2016}, \citeyear{Bilous2016}, 9 (27\%) pulsars have a cut off or flattening in the spectrum, according to observations at 110.4 and 135 MHz, 4 (12\%) pulsars (J0341+5711; J0408+551; J2337+6151; J2351+8533) the spectrum remains rising, and there is no estimate for 7 (21\%) pulsars (J0115+6325; J0141+6009; J0332+5434; J0343+5312; J1343+6634; J2312+6931; J2354+6155) the flux density is at 135 MHz, and therefore the behavior of the spectrum could not be estimated. Thus, excluding the 7 pulsars that do not have a 135 MHz flux density estimate, we find that of the 26 remaining pulsars, 22 (85\%) of the known pulsars found in the blind search exhibit signs of spectrum cut off or flattening.

For two pulsars (J0117+5914, J2229+6205), the average profiles from LPA1 observations could not be obtained, while the pulsars are visible in the stacked power spectra. Pulsar J2229+6205 was re-detected in a blind search using the LOFAR radio telescope (\citeauthor{Sanidas2019}, \citeyear{Sanidas2019}), while pulsar J0117+5914 was not detected in one-hour LOFAR observations.

The average profiles of known pulsars based on observations in one session are presented in Fig. \ref{Fig2}. Basically, the profiles have a simple triangular shape and their S/N at the peak is greater than 6. A part of pulsars are very weak (S/N$<6$), for example, J0341+5711, J0343+5312 and others. However, we are confident in the reliability of the results because all pulsars were detected in the records repeatedly and the maximum S/N in the profiles for different days correspond to close values of $P$ and $DM$. For some pulsars, the average profile was averaged using a moving average to increase the apparent S/N ratio.

	\section{Conclusion}

In the blind search for pulsars conducted on the LPA radio telescope, 35 pulsars were detected. The periods of the pulsars range from 0.15~s (PSR J0358+5413) to 1.93~s (PSR J0343+5312). The maximum dispersion measure was recorded for the pulsar J2302+6028 ($DM=156.7$ pc/cm$^3$). For the first time, estimates of the flux density at 110.4 MHz have been obtained for 33 pulsars, and the average profiles and their half-widths have been presented. It has been shown that up to 85\% of the pulsars found in blind searches have cut offs or flattenings in their integrated spectra at 110.4 MHz.

\section*{Acknowledgements}

The authors express their gratitude to L.B. Potapova for her assistance in preparing the figures and preparing LaTeX version of paper. We are also grateful to the anonymous referee for pointing out an error and making several suggestions that improved the paper. This work was carried out as part of a government assignment.

\end{document}